# Transient terahertz photoconductivity measurements of minority-carrier lifetime in tin sulfide thin films: Advanced metrology for an early-stage photovoltaic material

R. Jaramillo[1], Meng-Ju Sher[4], Benjamin K. Ofori-Okai[1], V. Steinmann[1], Chuanxi Yang[3], Katy Hartman[1], Keith A. Nelson[1], Aaron M. Lindenberg[2,4,5], Roy G. Gordon[3] & T. Buonassisi[1]

[1] Massachusetts Institute of Technology, Cambridge, MA 02139, USA
[2] Stanford Institute for Materials and Energy Sciences, SLAC National Accelerator Laboratory, Menlo Park, CA 94025, USA
[3] Department of Chemistry and Chemical Biology, Harvard University, Cambridge, MA 02138 USA
[4] Department of Materials Science and Engineering, Stanford University, Stanford, California 94305, USA
[5] PULSE Institute, SLAC National Accelerator Laboratory, Menlo Park, CA 94025, USA

**Abstract**

Materials research with a focus on enhancing the minority-carrier lifetime of the light-absorbing semiconductor is key to advancing solar energy technology for both early-stage and mature material platforms alike. Tin sulfide (SnS) is an absorber material with several clear advantages for manufacturing and deployment, but the record power conversion efficiency remains below 5%. We report measurements of bulk and interface minority-carrier recombination rates in SnS thin films using optical-pump, terahertz (THz)-probe transient photoconductivity (TPC) measurements. Post-growth thermal annealing in $H_2S$ gas increases the minority-carrier lifetime, and oxidation of the surface reduces the surface recombination velocity. However, the minority-carrier lifetime remains below 100 ps for all tested combinations of growth technique and post-growth processing. Significant improvement in SnS solar cell performance will hinge on finding and mitigating as-yet-unknown recombination-active defects. We describe in detail our methodology for TPC experiments, and we share our data analysis routines in the form freely-available software.

Solar cells need long-lived electronic excitations to deliver a useful amount of power. The minority-carrier lifetime of the light-absorbing semiconductor is a crucially important metric to evaluate an absorber material. This is firmly established for technologies based on Si wafers, for which lifetime measurements are routine. Minority-carrier lifetime is equally as important for thin-film solar cells based on materials other than Si, but measurements are not routine and are frequently inaccurate. Accurate lifetime metrology accelerates the technological development of materials and is a valuable screening metric for new materials.[1] Here we report measurements of minority-carrier lifetime in SnS thin films using optical-pump, THz-probe transient photoconductivity (TPC) measurements. SnS is an absorber with important advantages for solar cell manufacturing, but the record power conversion efficiency remains below 5%.[2] We demonstrate the impact of annealing and surface treatments on the bulk and surface minority-carrier recombination rates, and we illustrate the effect of minority-carrier recombination rates on solar cell performance. Throughout we highlight the challenges of measuring minority-carrier lifetime in thin films and we demonstrate methods to overcome these challenges.

In Section 1 we motivate the importance of minority-carrier lifetime and we introduce SnS as a solar cell absorber. In Section 2 we describe the sample set and procedures to control bulk and surface recombination rates. In Section 3 we present the results and compare to solar cell device performance. In Section 4 we discuss details of the measurement and data analysis. Section 5 contains discussion and conclusions.

**1. The importance of minority-carrier lifetime for solar cells. Introduction to SnS.**

Long minority-carrier lifetime is essential for high performance solar cells. The precise lifetime required to ensure high performance depends on the optical absorption length and excess-carrier mobility, and the minority-carrier lifetime by itself is a crude figure of merit. Nevertheless, a long lifetime is essential for high performance. In **Figure 1a** we compile reports of minority-carrier lifetime and solar cell efficiency for different thin film absorber materials: CdTe, Cu(In,Ga)(S,Se)$_2$ (CIGS), Cu$_2$ZnSnS$_4$ (CZTS), MAPbX$_3$ (methylammonium lead halide "perovskites"), InP, and SnS. We show only data for which both lifetime and device measurements were reported for samples that were synthesized in the same laboratory and using as close to the same procedure as is reasonably possible. Although it is possible to make a poorly performing solar cell from high lifetime material, the converse seems to be impossible. **Figure 1a** includes the lifetime measurements on SnS that compare directly to our previously published efficiency marks of 3.88% and 4.36% and that are described in Sections 2 and 3.[2,3]

All of the materials presented in **Figure 1a** have bandgaps appropriate for single-junction solar cells, can absorb the majority of above-bandgap light in 1 μm thick films, and are typically incorporated into solar cells as polycrystalline thin films. Within this set of technologically relevant constraints, the minority-carrier lifetime is a valuable figure of merit. The minority-carrier lifetime in silicon can exceed 100 μs for Czochralski-grown wafers, orders of magnitude higher than thin film materials.[4] However, this is compensated by the fact that the optical absorption depth in candidate thin film materials is orders of magnitude shorter than in silicon. The minority-carrier lifetime is too crude a metric to capture these differences.

For a more sophisticated solar cell figure of merit, we consider the dimensionless ratio ($F_{PV}$) of minority-carrier diffusion length to optical absorption length:

$$F_{PV} = \alpha\sqrt{D\tau} \qquad \textbf{Equation 1}$$

$D$ is the minority carrier diffusivity, $\tau$ is the bulk minority-carrier lifetime, and $\alpha$ is the optical absorption coefficient. In **Figure 1b** we present compiled data for $F_{PV}$ and solar cell efficiency, including results for wafer-based silicon technologies. With the exception of silicon, all of the materials represented strongly absorb light at energies above their respective band gaps. For each material we calculate $F_{PV}$ using $\alpha$ as measured at the knee the curve of $\log_{10}(\alpha(E))$. For the thin film materials we require that lifetime, diffusivity, absorption coefficient, and device measurements were reported for samples that were synthesized in the same laboratory and using as close to the same procedure as is reasonably possible. This requirement greatly reduces the number of data points compared to **Figure 1a**. For crystalline silicon we assume the values $D = 30$ cm$^2$s$^{-1}$ and $\alpha = 300$ cm$^{-1}$.

SnS is an absorber with several inherent advantages compared to materials that are widely used in solar cells, but its demonstrated efficiency is too low for commercial relevance. It is composed of non-toxic, Earth-abundant and inexpensive elements. SnS is an inert and water-insoluble semiconducting mineral (Herzenbergite) with an indirect bandgap of 1.1 eV, strong light absorption for photons with energy above 1.4 eV ($\alpha > 10^4$ cm$^{-1}$), and intrinsic *p*-type conductivity with carrier concentration in the range $10^{15} - 10^{17}$ cm$^{-3}$.[5–7] SnS evaporates congruently and is phase-stable up to 600 °C.[8,9] This means that SnS thin films can be deposited by thermal evaporation and its high-speed cousin, closed space sublimation (CSS), as is employed in the manufacture of CdTe solar cells. It also means that SnS phase control is simpler than for most thin film PV materials, notably including CIGS and CZTS. Therefore, efficiency stands as the primary barrier to commercialization of SnS solar cells. However this efficiency barrier cannot be overstated. The record efficiency for SnS solar cells is 4.36% (certified), and simulations indicate that the device performance is limited by minority-carrier lifetime.[10] For SnS solar cells to improve, research must focus on increasing the minority-carrier lifetime, and reliable lifetime metrology is indispensable.

**2. The SnS thin film sample set. Bulk annealing and surface passivation.**

Previous work has shown that SnS solar cells are improved by post-growth annealing and by surface oxidation.[2,3,11] Annealing in H$_2$S gas leads to improvements in short circuit current density ($J_{SC}$), fill factor (FF), and open circuit voltage ($V_{OC}$). Creating a thin SnO$_2$ layer in between the SnS absorber and the *n*-type buffer layer improves $V_{OC}$. These results suggest that H$_2$S annealing reduces the bulk defect density and improves the minority-carrier lifetime, and that surface oxidation passivates defects and reduces the surface recombination velocity (SRV). These hypotheses are supported by the observation that H$_2$S annealing promotes secondary grain growth, and that oxygen point defects at an SnS interface remove dangling bond states from the bandgap.[6,12] The solar cell studies have been performed on devices in the substrate configuration, so that the *p*-type absorber is deposited before the *n*-type buffer layer. The fabrication sequence is: SnS deposition, H$_2$S annealing, surface oxidation, buffer layer deposition.

We vary the annealing and surface oxidation treatments of SnS films grown by thermal evaporation and atomic layer deposition (ALD), and measure the effect on the bulk lifetime and the SRV. The films grown by thermal evaporation are 1000 nm thick and are grown using the same chamber and growth conditions as those used in the devices reported in refs. 3 and 13. The films grown by ALD are 380 nm thick and are grown using the same chamber and growth conditions are those used in the devices reported in refs. 2 and 13. The substrate temperature during growth is 240 and 200 ºC for thermal evaporation and ALD, respectively. The outstanding difference between the films studied here and the films used in devices is that these films are grown on fused quartz for compatibility with the THz absorption measurement, whereas for solar cells, the films are grown on Mo. The different annealing and oxidation treatments studied are summarized in **Table 1**.

## 3. Results: Bulk and interface minority-carrier recombination in SnS thin films, and comparison to solar cell performance.

In **Figure 2** we plot the minority-carrier lifetime ($\tau_0$) and SRV ($S$) for the full sample set. $\tau_0$ indicates the low-injection limit of the minority-carrier defect-assisted recombination lifetime (*c.f.* Section 4e). Sample TE3 corresponds to the solar cell devices made from thermally evaporated SnS and reported in refs. 3, 9, and 13. For this sample $\tau_0 = 38 \pm 1$ ps and the diffusion length $L_D = \sqrt{\tau_0 D_e} = 54.7 \pm 0.5$ nm, where $D_e$ is the minority-carrier diffusivity. This is in near-agreement with the value $L_D = 86 +/- 22/17$ nm that was estimated using a device model to fit quantum efficiency data measured on finished devices.[9] The trends in the results for $\tau_0$ and $S$ across our whole sample set support the hypothesis introduced above, that annealing in an $H_2S$ atmosphere increases $\tau_0$ and that oxidation of the surface decreases $S$.

Annealing in $H_2S$ increases $\tau_0$ for both thermally evaporated and ALD films. Samples TE3 and TE5 correspond closely to the process used to fabricate the 3.88% device reported in refs. 3 and 13. This so-called "baseline" process includes $H_2S$ annealing of the thermally evaporated SnS film, followed by oxidation in ambient air, followed by deposition of the Zn(O,S,N) buffer layer. Compared to the baseline process, sample TE4 was annealed for a longer time in higher $H_2S$ partial pressure and at higher temperature, resulting in a larger $\tau_0$. Sample TE2 was not annealed and has a smaller $\tau_0$ relative to the baseline process. The progression of increasing $\tau_0$ from samples TE2, to TE3, to TE4 shows the effect of $H_2S$ annealing. Similarly, $\tau_0$ increases going from non-annealed sample TE8 to annealed sample TE9, where both samples received the same $H_2O_2$ exposure. The ALD samples are also consistent with this trend, with the annealed sample (ALD2) having a larger $\tau_0$ than the non-annealed sample (ALD1). Annealing SnS films in $H_2S$ results in secondary grain growth, and we hypothesize that the increased minority-carrier lifetime results from a reduction in the density of extended crystallographic defects.[6]

Oxidation passivates the SnS surface and reduces $S$. This is seen by comparing samples TE1 and TE2. Sample TE2 was oxidized by exposure to ambient air for 24 hours, and $S = 4.5 \pm 0.3 \times 10^5$ cm s$^{-1}$. In contrast, the air exposure for sample TE1 was kept to a minimum: it was exposed to ambient for less than 10 s in total in between growth in high-vacuum ($< 10^{-7}$ Torr) and measurement in a medium-vacuum ($< 10^{-1}$ Torr). For this sample $S = (3.9 \pm 0.3) \times 10^6$ cm s$^{-1}$.

Samples ALD1 and 8 have the lowest SRV, with an average value $S = (8 \pm 1) \times 10^4$ cm s$^{-1}$. These samples were oxidized by H$_2$O$_2$ exposure immediately after growth, without intervening exposure to ambient air. We hypothesize that the reduction of $S$ by oxidation is due to the partial substitution of sulfur by oxygen, which is predicted to move dangling bond electronic states out of the bandgap.[12]

Although oxidation passivates the surface, an aggressive process is counterproductive. Sample TE7 was exposed to an O$_2$ plasma and the resulting SRV is extremely high. We hypothesize that ion bombardment in the O$_2$ plasma reactor creates more recombination-active defects than oxidation eliminates. H$_2$O$_2$ exposure of thermally evaporated films decreases $\tau_0$ and increases $S$. This is surprising, since the SnS solar cell efficiency record is held by a device grown by ALD and including surface passivation by H$_2$O$_2$, as with sample ALD2. However, the thermally evaporated films (TE8 and 6) were exposed to air before the H$_2$O$_2$ treatment, whereas the ALD-grown films (ALD1 and 8) were not (this is consistent with the fabrication procedure for the reported solar cell devices).[13] We hypothesize that the chemically aggressive H$_2$O$_2$ damages the previously formed surface oxide. The samples are polycrystalline, and certain physical processes may affect both $S$ and $\tau_0$. For example, a process that creates defects on the surface may also create defects on the internal grain boundaries. Such a process would increase $S$ and reduce $\tau_0$. It appears that H$_2$O$_2$ exposure of thermally evaporated, air-exposed films is such a process.

In **Figure 3a** we compare $\tau_0$ to the power conversion efficiency ($\eta$) of solar cell devices made from films with the same growth and processing procedures. We plot $\eta$ as measured in our laboratory, and the errorbars indicate the distribution of measurements on multiple devices. For these measurements light-masking was not used, and the spectrum of the solar simulator was not frequently calibrated. Therefore, these results are less accurate than the certified data reported in refs. 2 and 3 and reproduced in **Figure 1**. We observe a slight rising trend in $\eta$ with $\tau_0$. Relatively small changes in $\tau_0$ are difficult to correlate to device performance because there are many factors that affect $\eta$: as **Figure 1** demonstrates, a large $\tau_0$ is a necessary but not sufficient condition for a high performance solar cell.

The open-circuit voltage ($V_{OC}$) is rather directly affected by surface recombination. The recombination current density is larger at $V_{OC}$ than at any other point in the power-generating quadrant of the current-voltage curve. Furthermore, at $V_{OC}$ the absorber is close to flat-band conditions. As a result the product ($pn$) of majority and minority-carrier concentrations is enhanced at the interface. In **Figure 3b** we compare $S$ and $V_{OC}$. There is a clear trend of decreasing $V_{OC}$ with increasing $S$.

$\tau_0$ and $S$ were measured with optical-pump, THz-probe TPC. Each sample was measured with at least two combinations of pump wavelength and fluence in order to increase the confidence in the estimated bulk and surface recombination rates. Several samples were measured repeatedly in two different laboratories. $\tau_0$ and $S$ were estimated for each sample by global fits to multiple data sets. In **Figure 4** we show representative data and fits for a single sample (TE4). The experiments and data analysis are described in detail in **Section 4**.

## 4. Transient photoconductivity measurements of minority-carrier lifetime in thin films

Minority-carrier lifetime is determined by measuring the excess concentration of minority-carriers that results from optical excitation.[14,15] By "excess" we mean a state away from thermal equilibrium. The excitation and measurement can be steady-state or transient. Unfortunately the quantity measured is not the excess minority-carrier concentration, but a function of the same. Photoluminescence measures the rate of optical recombination, and photoconductivity (or, equivalently, free carrier absorption) measures the electrical conductivity; both are functions of the excess majority and minority-carrier concentrations. The spatial profile of excess-carriers through the sample is rarely measured but can be modeled with assumptions. For wafer-based materials such as Si the experimental methodologies and the assumptions involved are mature, and the results are often accurate.[14,16] For thin films these same methodologies and assumptions are usually inappropriate. The parameter space for measurements of thin films is very different than for wafers, and for most thin film materials the unknowns are more numerous and problematic.

Here we discuss in detail seven aspects of our experimental procedure and data analysis:

a) Key assumptions: defect-assisted recombination, and absence of electric fields
b) Optical-pump, THz-probe transient photoconductivity measurements: Description of the experiments
c) Optical-pump, THz-probe transient photoconductivity measurements: Data pre-processing
d) The spatial profile of excess-carriers
e) Measurements under high-injection conditions
f) The challenge of unknown diffusivity
g) Sample heating and other phenomena that affect the data at long times

These details arise in the case of optical-pump, THz-probe measurements on SnS thin films, but the methods discussed may apply to a wide range of materials and to related techniques.

*4a. Key assumptions: defect-assisted recombination, and absence of electric fields*

We assume that the minority-carrier lifetime is limited by defect-assisted recombination, and that the competing processes radiative recombination and Auger recombination do not affect the lifetime. This is applicable to the majority of thin film solar cells. Optical recombination is dominant only for solar cells based on epitaxial single crystals of direct bandgap semiconductors such as GaAs, which are not discussed here.[17] Auger recombination is dominant for solar cells that operate at very high injection levels or have high bulk doping densities, such as certain types based on monocrystalline Si, but is not relevant here.[18] The Auger and defect-assisted recombination rates are equal for injection level $n_A = \sqrt{1/\Gamma\tau}$, where $\Gamma$ and $\tau$ are the Auger coefficient and defect-assisted recombination lifetime, respectively. For $n \ll n_A$ Auger recombination is irrelevant. $\Gamma$ is unknown for SnS so we use the value for Si, $\Gamma = 1 \times 10^{30}$ cm$^6$ s$^{-1}$. For the values of $\tau$ reported here $n_{eq} > 10^{20}$ cm$^{-3}$ and $n \ll n_A$ is satisfied for all of the data.

We assume that there are no electric fields in the film, either in equilibrium or under illumination. Our samples are bare thin films grown on fused quartz and are capped by ultra-thin (< 5 nm) oxide layers, so it is reasonable to assume that there are no heterojunctions and associated electric fields. We assume that there is no surface voltage due to trapped charge at the surface. Finally, we assume that charge moves by ambipolar diffusion such that no space charge develops during the measurement. This assumption may be unjustified in some cases. However, it is essential in order to restrict the model to diffusive dynamics, instead of coupled drift-diffusion and electrostatics. This makes the problem tractable for curve-fitting with a personal computer and widely available software. The accuracy of assuming ambipolar diffusion may be assessed by comparing the dielectric relaxation time to the minority-carrier lifetime. The relaxation time $\tau_r$ is equal to $\varepsilon_0\varepsilon_r/\sigma$ where $\varepsilon_0\varepsilon_r$ and $\sigma$ are the static dielectric susceptibility and electrical conductivity, respectively.[19] For short times $t < \tau_r$ ($t$ is the time elapsed since the optical pump) electrons and holes diffuse independently with different diffusion coefficients. The resulting accumulation of space-charge results in an electric field and a Dember potential. For long times $t > \tau_r$ the space-charge is screened and electrons and holes diffuse together with the ambipolar diffusion coefficient. $\tau_r$ is difficult to estimate for SnS. $\varepsilon_r$ is anisotropic and has not been measured at low frequency. Density functional theory predicts that $\varepsilon_r$ ranges from 34 – 52 depending on orientation.[20] The DC conductivity that we measure is affected by grain-boundary scattering, whereas $\tau_r$ is controlled by the intra-granular mobility. Measurements of mobility in single-crystal SnS range from 10 to well over 1000 cm$^2$ V$^{-1}$ s$^{-1}$.[21,22] Our own THz spectroscopic measurements of the complex dielectric response of a bare SnS film at equilibrium find $\sigma = 1.0 \pm 0.8$ S and $\varepsilon_r = 29 \pm 1$ at 1 THz, in reasonable agreement with the aforementioned published results. Using these values we estimate that $\tau_r = 1 - 10$ ps. This implies that ambipolar diffusion is an accurate assumption for all but the initial time points in our experiments, and that the excess-carrier dynamics should be little affected by electric fields due to different hole and electron diffusivities.

*4b. Optical-pump, THz-probe transient photoconductivity measurements: Description of the experiments*

In a TPC experiment excess-carriers are generated by a short pulse of above-bandgap light, and the relaxation to equilibrium is measured by the absorption of long-wavelength light. The optical pump and the THz probe are generated from an amplified femtosecond laser source and are separated in time by a delay line. Two different optical pump wavelengths ($\lambda = 400$ and 800 nm) are used in this experiment to generate excess-carriers at different depths. The THz signal is measured in transmission through the film-on-substrate sample. The bandwidth of the THz measurement is approximately 2 THz. The experimental setup is described fully in ref. 23.

For the TPC experiments we measure the amplitude of the transmitted electric field ($T$) at a fixed position in the THz waveform as a function of pump-probe delay time ($t$), and compare this signal to the amplitude transmission ($T_0$) in the absence of a pump. $T_0$ is measured at least 1 ms after the most recent optical pump. The raw data is $(T(t) - T_0)/T_0 \equiv \Delta T/T_0$. The amplitude transmission is measured at the peak of the THz waveform. We have explicitly confirmed that there is negligible phase change due to the presence of the SnS film. This is consistent with our time-domain THz spectroscopy measurements of a film at equilibrium that show that the "crossover frequency" $f_c > 2$ THz, or equivalently that the Drude scattering time is less than 80 fs. Therefore the complex conductivity of the film at the THz frequencies used should be predominantly real.[24]

The sample is held either in air, in a box purged with nitrogen gas, or in a cryostat under vacuum, depending on the desired amount of air-exposure. All measurements are performed at room temperature. The spot size of the optical pump is between 2 – 3 mm (radius at 1/e intensity), and the THz probe is between 0.7 – 0.8 mm. We assume that the excess-carrier concentration is laterally uniform as probed by the THz beam. The pump time envelope is 100 – 200 fs (full-width at half-maximum), and the THz envelope is 500 – 700 fs. For transient measurements we do not analyze the THz waveform, and therefore the time resolution is comparable to the THz envelope. We account for finite time resolution by modeling the pump as a Gaussian in time, and fitting the initial rise of the data to determine the width of the pulse. For all data reported here the width is 500 – 700 fs (standard width). The fluence ($N_0$) of the pump pulse is measured independently and is not a free parameter in the fits. $N_0$ is the time-integrated fluence for a single pulse in units of particles/area. The pump area is calculated for a circle with radius equal to the experimentally-measured radius at which the intensity falls by a factor of 1/e from the maximum. For all data reported here $N_0$ is between $2 \times 10^{13}$ and $6 \times 10^{14}$ cm$^{-2}$. Reflection from the front surface is measured independently and accounted for. The spatial generation profile through the film is determined by the optical absorption coefficient ($\alpha$) which we measured independently: $\alpha(\lambda = 400$ nm$) = 7.27 \times 10^5$ cm$^{-1}$, $\alpha(\lambda = 800$ nm$) = 4.84 \times 10^4$ cm$^{-1}$. Due to the large optical absorption coefficient, reflection at the rear surface does not make a significant contribution to the generation profile.

*4c. Optical-pump, THz-probe transient photoconductivity measurements: Data pre-processing*

In order to extract useful semiconductor device parameters, $\Delta T/T_0$ needs to be converted into excess-carrier concentration. Our films are at most 1 μm thick and the optical wavelength at 1 THz is approximately 300 μm. Therefore we use the equation for the amplitude transmission through an air / thin film / substrate structure:[25]

$$T_{TF} = \frac{2}{1 + n_S + Z_0 \sigma d} \quad \quad \textbf{Equation 2}$$

$n_S$ is the substrate index of refraction, $Z_0$ is the impedance of free space, $\sigma$ is the film conductivity at THz frequencies, and $d$ is the film thickness. Our substrates are fused quartz with $n_S = 2.13 \pm 0.01$ at 1 THz as measured by THz spectroscopy on bare substrates. The optical pump changes $\sigma$, and the other terms in Eq. 2 remain the same. The amplitude transmission to air at the back of the substrate cancels in the ratio $\Delta T/T_0$, and due to the time-domain nature of the experiment we can ignore multiple reflections within the substrate. We can therefore relate $\Delta T/T_0$ to the relevant material parameters:

$$\frac{-\Delta T/T_0}{1+\Delta T/T_0} = \frac{\sigma_1/\sigma_0}{1+\dfrac{1+n_S}{Z_0 d\sigma_0}} = \left(\frac{1}{1+\dfrac{1+n_S}{Z_0 d\sigma_0}}\right)\left(\frac{m_h^*}{m_e^*}+1\right)\frac{\bar{n}(t)}{p_0} \qquad \textbf{Equation 3}$$

These equations are written for a $p$-type semiconductor. The film conductivity $\sigma(t) = \sigma_0 + \sigma_1$, where $\sigma_0$ is the equilibrium value and $\sigma_1$ is the excess photoconductivity. $m_h^*$ and $m_e^*$ are the hole and electron effective masses, respectively. $\bar{n}(t)$ is the excess minority-carrier concentration (the bar indicates an average through the film) and $p_0$ is the equilibrium majority carrier concentration. The second equality follows from the expression $\sigma_1 = q\mu_e \bar{n} + q\mu_h \bar{p}_1$ and the assumptions that $\bar{n} = \bar{p}_1$ and that $\mu_e/\mu_h = m_h^*/m_e^*$. $\bar{p}_1$ is the excess majority carrier concentration and $\mu_e$ and $\mu_h$ are the electron and hole mobility, respectively. In the absence of published measurements of the effective masses we take $m_h^*/m_e^* = 2$, the average of $m_h^*/m_e^*$ along the crystal axes as calculated by density functional theory.[26] We assume that the mobilities are constant in time. This is valid because the peak injection levels are always below the level of degeneracy, and because the timescales involved are much longer than typical hot carrier relaxation times. With these assumptions we use **Eq. 3** to calculate $\bar{n}(t)$ from $\Delta T/T_0$.

*4d. The spatial profile of excess-carriers*

Low-frequency probes such as IR, THz, and DC measure the average conductivity through the film, and the terms that appear in **Eq. 2-3** are averages through the film. However, the spatial distribution of excess-carriers is rarely uniform, and due to competing processes the instantaneous decay rate $r(t) = \dfrac{d\log(\bar{n})}{d\bar{n}}$ depends on the spatial distribution. The spatial distribution approaches a flat line (*i.e.* becomes independent of position) only when the minority-carrier lifetime is much longer than the diffusion time and the surfaces are not recombination-active. In order to accurately estimate the recombination rates the excess-carrier spatial profile must be considered. Here we discuss a model for the spatial distribution of excess-carriers, and we explore the parameter spaces within which the spatial distribution must be explicitly considered in order to interpret experiments.

The dynamics of excess-carriers are governed by diffusion, drift, generation, and recombination. We assume spatial uniformity in the plane of the sample, so that all quantities vary only along the out-of-plane axis ($x$). As described above we neglect drift currents. Therefore the dynamics of excess minority-carriers are governed by the diffusion equation with generation and recombination as shown in **Equation 4**:

$$\frac{\partial n}{\partial t} = D\frac{\partial^2 n}{\partial x^2} + G - \frac{n}{\tau}$$

$$\left.\frac{\partial n}{\partial x}\right|_{x=-d/2} = Sn(x=-d/2)$$

$$\left.\frac{\partial n}{\partial x}\right|_{x=d/2} = -Sn(x=d/2)$$

**Equation 4**

$n(x, t)$ is the excess minority-carrier concentration as a function of space and time, $D$ is the diffusivity, $\tau$ is the bulk minority-carrier lifetime, $G$ is the generation, and $S$ is the SRV. The film thickness is $d$, and the interfaces are at $x = +/- d/2$. The film is illuminated from the front surface ($x = -d/2$). We assume that the SRV is the same at front and back surfaces. Although this may be inaccurate in some cases, recombination at the back surface is not relevant in our samples. The geometry is illustrated in **Figure 4**.

Solutions to **Eq. 4** have been discussed extensively in the context of TPC measurements of semiconductor wafers.[16,27] Solutions have the form

$$n = e^{-t/\tau}\sum_j \left(A_j e^{-\alpha_j^2 Dt}\cos(\alpha_j x) + B_j e^{-\beta_j^2 Dt}\sin(\beta_j x)\right)$$

**Equation 5**

The frequencies $\alpha_n$ and $\beta_n$ depend on $D$, $S$, and $d$, and the amplitudes $A_j$ and $B_j$ depend on the spatial distribution of generation. When calculating the average $\bar{n}(t)$ the sine term disappears. The surface recombination rates $\alpha_j^2 D \equiv r_{s,j} \equiv 1/\tau_{s,j}$ characterize the process of carriers diffusing to the surfaces and recombining there. The frequencies satisfy $\alpha_1 < \alpha_2 < \alpha_3 < \ldots$ At short times several frequencies contribute to the measured signal, and $\bar{n}(t)$ exhibits a multi-exponential decay. For times $t \gg \tau_{s,2}$ only the fundamental frequency contributes, and $\bar{n}(t)$ exhibits single-exponential decay with effective lifetime $\tau_{eff}$:[27]

$$\frac{1}{\tau_{eff}} = \frac{1}{\tau} + \alpha_1^2 D \approx \frac{1}{\tau} + \left(\frac{d^2}{\pi^2 D} + \frac{d}{2S}\right)^{-1}$$

**Equation 6**

Experiments on wafers of highly developed materials usually extend to $t \gg \tau_{s,2}$, in which case $\tau_{eff}$ is easily determined by a single exponential fit to the data. In this case the spatial distribution of excess-carriers does not affect the measurement of $\tau_{eff}$. However, this condition may not be met for thin films.

The surface recombination times $\tau_{s,j}$ satisfy $\tau_{s,j} < t_d/4\pi^2$ for all $j > 1$, where $t_d = d^2/D$ is the diffusion time. Therefore, the TPC data must extend to times $t \gg t_d/4\pi^2$ in order for $\tau_{eff}$ to be determined by a single exponential fit to the data. For an ideal experiment with arbitrary sensitivity this could always be satisfied. However, real experiments have finite sensitivity, and the requirement of collecting data for $t >> t_d/4\pi^2$ may not be met. The total signal decays no faster than $\tau_{eff}^{-1}$. Therefore, experiments are sensitive at times $t << \tau_{eff}$, and insensitive at times $t >> \tau_{eff}$. By combining the known solutions to Eq. 3 with this estimate of experimental sensitivity we identify parameter spaces within which the spatial distribution of excess-carriers can and cannot be ignored when fitting TPC data. For $\tau_{eff} >> t_d/4\pi^2$ the fundamental mode is easily measured, and the spatial distribution of excess-carriers does not affect the estimation of $\tau_{eff}$. For $\tau_{eff} << t_d/4\pi^2$ the data is not well-characterized by a single exponential decay, and the spatial distribution of excess-carriers must be considered in order to estimate $\tau_{eff}$. In **Figure 5** we show these parameter spaces in the $D$-$\tau$ plane for discrete values of $S$ and fixed film thickness $d = 1$ μm, which is typical for thin film solar cell absorbers. We also indicate parameter regimes that characterize different absorber materials. For long minority-carrier lifetimes and/or fast diffusion, the spatial distribution does not affect the determination of $\tau_{eff}$. For short minority-carrier lifetimes and /or slow diffusion, the spatial distribution must be accounted for in order to determine $\tau_{eff}$.

In **Figure 6** we show calculated $\bar{n}(t)$ and $n(x, t)$ for two representative cases. **Figure 6a** shows the case of a long minority-carrier lifetime ($\tau$ = 125 ns) and a small diffusivity ($D$ = 0.15 cm$^2$/s), typical of MAPbX$_3$.[28–32] **Figure 6b** shows the case of a short minority-carrier lifetime ($\tau$ = 0.03 ns) and a larger diffusivity ($D$ = 1 cm$^2$/s), typical of SnS. For the long-lifetime material, the higher-order frequencies $j > 1$ decay quickly relative to $\tau_{eff}$, and $\tau_{eff}$ can be estimated easily from an exponential fit to the data at long times $t >> d^2/4\pi^2 D$. The spatial distribution of excess-carriers approaches a cosine half-cycle and the shape becomes fixed in time. In contrast, for the short-lifetime material, the higher order frequencies do not decay quickly relative to $\tau_{eff}$. The spatial distribution of excess-carriers changes continually and does not reach the half-cosine shape within a typical experimental time window. In this case the spatial distribution must be explicitly modeled in order to estimate $\tau_{eff}$.

$\tau_{eff}$ depends on both bulk and surface recombination. The exponential decay at long times reflects the minority-carrier lifetime ($\tau$) only for the case of small $\tau$ or very well pacified surfaces ($S < 100$ cm/s). In **Figure 7** we plot the ratio $\tau_{eff}/\tau$ as a function of $D$ and $\tau$ for discrete values of $S$ and for a fixed film thickness $d = 1$ μm. For wafer-based materials such as Si, there are well-known techniques for passivating the surface ($S < 1$ cm/s can be achieved by immersion in HF) in order to directly measure $\tau$.[14] For thin film materials, this degree of control over surface recombination has not been achieved, and as a result $\tau_{eff}$ is usually dependent on $\tau$, $d$, $D$, and $S$.

*4e. Measurements under high-injection conditions*

Most solar cells operate at low injection conditions, meaning that the excess-carrier concentration under solar illumination remains lower than the equilibrium majority carrier concentration. Due to limited experimental sensitivity many minority-carrier lifetime measurements are performed under high-injection conditions. Minority-carrier recombination rates depend on the injection level due to the statistics of trap occupation.[33] Under the simple assumption of a mid-gap trap state, the minority-carrier lifetime ($\tau$) can be written as:

$$\tau(n) = \tau_0 \frac{1 + 2(n/p_0)}{1 + (n/p_0)} \qquad \textbf{Equation 7}$$

$\tau_0$ is the minority-carrier lifetime at low injection. **Eq. 7** shows that the minority-carrier lifetime can be over-estimated by a factor of two if measurements are performed under high-injection without appropriate corrections.

In order to account for varying injection conditions we use **Eq. 7** to model the effect of $n$ on $\tau$. This means that **Eq. 4** becomes a nonlinear differential equation and we need to solve it numerically instead of using the linear, analytic solutions. In **Figure 8** we compare the model with and without including the injection-dependence of the minority-carrier lifetime using data and parameters representative of sample (TE4). The injection-dependence has a substantial effect on the model for $\bar{n}(t)$ and on the resulting estimate for $\tau_0$.

**Eq. 7** expresses a basic assumption about the nature of the dominant recombination-active defect. We assume that the defect state is near the middle of the bandgap, and that it can assume only two charge states. We assume that the semiconductor is either *n*- or *p*-type, not intrinsic. Finally, we assume that the recombination lifetime at low injection is the same ($\tau_0$) for both minority and majority carriers. These assumptions are necessary for SnS, for which little is known about the recombination-active defects. For better characterized materials **Eq. 7** could be easily refined without changing the overall approach.

Surface recombination must also depend on injection conditions because it is also governed by the statistics of trap occupation.[34] However, this is important only for materials with well-passivated surfaces. For example, $S < 1000$ is readily achieved at the Si / SiO$_2$ interface, and the excess-carrier concentration at the surface can be large. In this situation the injection-dependence of $S$ can significantly affect lifetime measurements and solar cell operation.[34,35] However, for most thin films $S$ is large, and the excess-carrier concentration at the surface is small. For our measurements we estimate that $n(x = -d/2) < p_0$ holds for all data except at very short times after photoexcitation. Injection-dependence of $S$ would result in a nonlinear dependence of the peak measured injection ($\bar{n}_{MAX}$) on the total pump fluence $N_0$. We find that $\bar{n}_{MAX}$ is weakly nonlinear for over two orders of magnitude variation in $N_0$, from $10^{13}$ to $10^{15}$ cm$^{-2}$. Therefore, it is appropriate to assume a constant value for $S$.

*4f. The challenge of unknown diffusivity*

For wafer-based materials such as Si the majority and minority-carrier diffusivity are well characterized and can be treated as known quantities for TPC analysis. This is not the case for thin films. For our films we only know the in-plane majority carrier Hall mobility, and we rely on assumptions to relate this to required quantities: the in-plane THz-frequency mobilities that appear implicitly in **Eq. 2**, and the out-of-plane diffusivity that appears in **Eq. 4**. We use three tactics to handle these unknown quantities. First, we assume that all diffusion is ambipolar so that the diffusivity $D$ in **Eq. 4** depends on the excess-carrier concentration, the effective masses, and the majority carrier mobility. Second, we assume that the in-plane mobilities that factor into $\sigma$ in **Eq. 2** are equal to the out-of-plane mobilities that relate to the diffusivity $D$ in **Eq. 4**. Finally, we use a global fitting routine to simultaneously optimize the model to multiple data sets.

The assumption that there are no electric fields in the film requires that excess majority and minority-carriers diffuse together with an ambipolar diffusivity $D_a$:[36]

$$D_a = \frac{n+p}{n/D_e + p/D_h} = D_e \frac{p_0 + 2n}{p_0 + n\left(1 + m_h^*/m_e^*\right)} \quad \textbf{Equation 8}$$

$D_h$ and $D_e$ are the bare hole and electron diffusivities and are related to the mobilities by the Einstein relation $D_{h,e} = k_B T \mu_{h,e}/q$. $D_a$ is injection-dependent and introduces another term nonlinear in $n$ to **Eq. 4**. The effect of including this nonlinearity is illustrated in **Figure 8** where we compare the model with and without the injection-dependence of $D_a$ using data and parameters for a representative sample (TE4). For SnS the injection-dependence of $D_a$ is not strong, and its effect on the model for $\bar{n}(t)$ is negligible.

The effects of diffusion and surface recombination on $\bar{n}(t)$ are correlated. The diffusivity controls how quickly carriers move towards a surface, and the SRV controls how quickly they recombine once they arrive at the surface. The covariance between $D$ and $S$ becomes extremely large if both are allowed to vary while fitting **Eq. 4** to experimental data, and it is necessary to impose additional constraints in order to draw useful conclusions from the data. To this end, we constrain the in-plane mobility at THz frequencies to be related by the Einstein relation to the out-of-plane diffusivity. In this way the parameter $D$ affects the overall scale of the $\bar{n}(t)$ as can be seen by rewriting **Eq. 3**:

$$\bar{n} = \left(\frac{-\Delta T/T_0}{1+\Delta T/T_0}\right)\left(1 + \frac{1+n_S}{Z_0 d\sigma_0}\right)\left(\frac{1}{m_h^*/m_e^* + 1}\right) p_0 \approx \left(\frac{-\Delta T/T_0}{1+\Delta T/T_0}\right)\left(\frac{1}{m_h^*/m_e^* + 1}\right)\frac{1+n_S}{Z_0 dq}\frac{1}{\mu_h} \quad \textbf{Equation 9}$$

The approximation follows from the fact that $Z_0 d\sigma_0 = 0.0002 - 0.02$ for 1 μm thick films with $\sigma_0 = 0.005 - 0.5$ S/cm, which covers our entire sample set. Combined with the Einstein relation and the assumption $\mu_e/\mu_h = m_h^*/m_e^*$, **Eq. 9** shows that the fitting parameter $D_e$ controls the scale of the data $\bar{n}(t)$. This serves as an additional constraint on the free parameters and reduces the covariance between $D$ and $S$. The approximation in **Eq. 9** helps to visualize the relationships between the parameters, but is not necessary to apply this constraint.

The assumption that in-plane and out-of-plane mobilities are equal is peculiar for a layered, non-cubic material such as SnS. However, X-ray diffraction data shows that our films grown on glass or $SiO_2$ have a distribution of orientations, and are not simply oriented along the long axis.[37] The area measured by the TPC experiments is larger than 0.1 cm$^2$ and the individual crystal grains have lateral dimension on the order of 100 nm. The measurement averages over billions of crystal grains, thereby justifying the assumption of equal in-plane and out-of-plane mobility on this scale. For more uniformly oriented samples additional assumptions could be used to relate the in-plane THz-frequency mobility to the out-of-plane diffusivity.

Finally, we use a global fitting routine to further constrain the free parameters and to reduce covariance between $D$ and $S$. By fitting simultaneously to multiple data sets we can constrain certain parameters to be constant across different data sets. For a given sample we hold $S$, $\tau_0$, and $D_e$ constant while fitting to data sets with varying the pump wavelength and fluence. For samples that were measured repeatedly in two different laboratories, both sets of measurements were included in the same global fit.

*4g. Sample heating and other phenomena that affect the data at long times*

At long times, the data $\Delta T/T_0$ is expected to decay exponentially to zero with time constant $\tau_{\text{eff}}$. However, our data instead saturates at a finite value $(\Delta T/T_0)_{\text{SAT}}$. In **Figure 9** we show representative data for a number of different samples with a fixed pump (400 nm, 11 µJ/cm$^2$), and in the inset we show the dependence of -$(\Delta T/T_0)_{\text{SAT}}$ on pump fluence for a single sample. $(\Delta T/T_0)_{\text{SAT}}$ varies slightly between samples, and increases monotonically with pump fluence.

We hypothesize that $(\Delta T/T_0)_{\text{SAT}}$ does not reflect the processes of excess-carrier recombination that we address here, and does not depend in a straightforward way on $\tau_0$ and $S$. There are two sets of evidence to support this hypothesis. One is the observation that the model described above does not fit the data well unless $(\Delta T/T_0)_{\text{SAT}}$ is explicitly removed or separately accounted for. This is particularly true for global fits to multiple data sets with varying pump wavelength and fluence. At long times the model predicts an exponential decay with time constant $\tau_{\text{eff}}$. If allowed to fit the raw data including $(\Delta T/T_0)_{\text{SAT}}$ the model finds unrealistically large values of $\tau_0$, well in excess of 1 ns, that depend strongly on the pump. $(\Delta T/T_0)_{\text{SAT}}$ may reflect long-lived electronic excitations in the SnS film such as slow carrier de-trapping at grain boundaries or shallow defects, and such phenomena would affect solar cell device performance by limiting the carrier mobility. However, such phenomena are not directly related to the minority-carrier lifetime $\tau_0$ for defect-assisted recombination of minority-carriers in the conduction band.

The second line of evidence in support of our hypothesis that non-zero $(\Delta T/T_0)_{\text{SAT}}$ does not depend on $\tau_0$ and $S$ is that transient heating can plausibly account for the observations. Below we discuss one such mechanism, the direct heating of the film by the laser pulse. We do not propose this effect as the clear explanation for the observed $(\Delta T/T_0)_{\text{SAT}}$. Rather, we use it to illustrate the existence of transient processes with widely varying time scales, and the challenge of addressing these processes with a limited experimental time window.

A short laser pulse with energy in the range $10^{-5} - 10^{-4}$ J/cm$^2$ can raise the temperature of a 1 µm SnS film by between 0.1 and 1 K. This can result in a thermal response in $\Delta T/T_0$ on the order of $10^{-3}$. Crucially, the timescale on which the film cools to ambient temperature is longer than the timescale for excess-carrier recombination, but is shorter than the experimental repetition rate.

Therefore, this thermal effect can be easily mistaken for a long-lived electronic excitation, with characteristic lifetime exceeding the experimental window.

We demonstrate this effect by solving a thermal model of the air / film / substrate system, using parameters appropriate to SnS and fused quartz where available. The SnS heat capacity ($C$), density ($\rho$) and thermal conductivity ($K$) are 45 J mol$^{-1}$ K$^{-1}$, 5.08 g cm$^{-3}$, and 1.3 W m$^{-1}$ K$^{-1}$, respectively.[38,39] We hold both the air and the substrate at ambient temperature ($\Theta_A$). We take the air / film thermal boundary conductance to be $5 \times 10^{-4}$ W cm$^{-2}$ K$^{-1}$, appropriate for a surface with emissivity of unity.[40] We take the film / substrate thermal boundary conductance to be $5 \times 10^{3}$ W cm$^{-2}$ K$^{-1}$, as for the Si / SiO$_2$ interface.[41] We assume that the electronic excitation following the optical pump quickly decays into heat, so that the initial temperature distribution at $t = 0$ is $\theta_A + (\alpha Q_0 / \rho C) e^{-\alpha(x-d/2)}$ where $\alpha$ is the optical absorption coefficient and $Q_0$ is the pump energy density. We plot in **Figure 10a** the resulting temperature profile through the film as a function of time. For $t < 1$ μs there is some re-distribution of heat through the film, but little loss to the substrate. The film starts cooling at times $t > 1$ μs. This is simply the thermal diffusion time ($t_\Theta$) for $d = 1$ μm of SnS: the Kelvin diffusivity is $\kappa = K/\rho c = 0.0086$ cm$^2$ s$^{-1}$, and $t_\Theta = d^2/\kappa = 1$ μs. For times $t < t_\Theta$ the thickness-averaged sample temperature ($\Theta_S$) is nearly equal to $\Theta_A + 0.2$ K.

We measured directly the effect of changing temperature on the THz signal. We measured the full THz transmitted waveform through sample TE3 while varying $\Theta_S$ using a heating stage and with no pump beam. The amplitude transmission decreased with increasing $\Theta_S$, with no noticeable shift in phase. We use the resulting data to parameterize $\Delta T_\Theta / T_0$ as a function of $\Theta_S$, where $\Delta T_\Theta = T(\Theta_S) - T(\Theta_A)$, and $T_0 = T(\Theta_S = \Theta_A)$. This can be combined with the output of thermal simulations to produce an estimate of the signal $\Delta T_\Theta / T_0$ as a function of $Q_0$. In **Figure 10b** we show an example of the simulated $\Delta T_\Theta / T_0$ as a function of time for typical experimental parameters. To this we add the TPC signal $\Delta T_{TPC}/T_0$ calculated using the model described above and material parameters taken from sample TE3. For a material such as SnS with a short minority-carrier lifetime, the timescales for excess-carrier recombination and sample cooling are well-separated, making these effects easier to distinguish. For longer-lifetime materials these timescale may become comparable, which would complicate the analysis. The analysis of thermal effects if made difficult by the fact the timescale for thermal transients is order of magnitude longer than the time that can be covered by typical delay stages.

Transient heating of the thin film as described above accounts for only a fraction of the measured $(\Delta T/T_0)_{SAT}$. We hypothesize that other thermal transients may fully account for these observations. This hypothesis can be tested by changing the sequence of visible and THz pulses and by changing the thermal conductivity of the sample environment. For sample TE1 that was measured in vacuum, we see evidence of extended thermal transients on a timescale of minutes. This is consistent with an increase in the thermal time constant of the sample, substrate and sample holder due to the lack of convective cooling in vacuum. On a practical level, this can be dealt with by increasing the experimental repetition rate so that the sample remains heated for the entire measurement sequence. Measurements extending over much longer time delays would enable $(\Delta T/T_0)_{SAT}$ to be studied directly and better understood. Unfortunately this may require that the pump-probe time delay extend to the experimental repetition period of 1 ms or longer, whereas typical delay stages have a maximum extent on the order of 1 ns.

We account for $(\Delta T/T_0)_{SAT}$ by adding to the model a term $\bar{n}_f(t) = f \int_{t_1}^{t} dt' N(t')$, where $f$ is a fit parameter and $N(t)$ is the time-dependence of the pump fluence such that the total fluence $N_0 = f \int_{t_1}^{t_2} dt' N(t')$. The integration limits $t_1$ and $t_2$ cover the full extent of a single pulse. $\bar{n}_f(t)$ expresses our assumption that $(\Delta T/T_0)_{SAT}$ increases monotonically with pump fluence and results from processes with an asymmetrical time response. The signal $(\Delta T/T_0)_{SAT}$ rises as quickly as the pump arrives but decays on a time scale much longer than the experimental time window. The $f$ parameter is relatively constant across our entire sample set for a fixed set of pump wavelength, total fluence, and measurement laboratory. The addition of this $f$ parameter to the model is an unfortunate necessity and reflects our incomplete understanding of the electrical and thermal response of our samples and the experimental apparatus. The hypothesis that $(\Delta T/T_0)_{SAT}$ results from long-lived electronic excitations in the sample such as surface traps could be tested by using varying temperature or an infrared bias light to de-trap charges during the TPC measurements.

## 5. Discussion and conclusion

**Figure 1a** makes clear that $\tau_0$ of SnS thin films must increase by one hundred-fold in order for SnS solar cells to join CdTe, CIGS and MAPbX$_3$ as a high-performing thin film solar cell technology. Thermal annealing in H$_2$S gas helps, but we only achieve increases by a factor of two for the conditions studied here. It remains to be seen whether more intensive annealing conditions (higher H$_2$S pressure, higher temperature, longer time) will yield further increases in $\tau_0$. Elsewhere we have reported evidence that $\tau_0$ increases upon raising the film growth substrate temperature, and this is another promising avenue for research.[9] Unfortunately, growth and annealing at higher temperature are made difficult by the high vapor pressure of solid SnS. CSS allows the deposition of high vapor pressure materials at high substrate temperatures, and is an enabling process for the manufacture of CdTe thin film solar cells. We suggest that growth of SnS tin films by CSS at substrate temperatures above 300 °C, and perhaps in H$_2$S carrier gas, may yield films with significantly higher $\tau_0$.

By varying the growth technique and the post-growth treatment we have attempted to control intrinsic recombination-active defects such as grain boundaries, extended intra-granular crystallographic defects, and intrinsic point defects. However, the fact that the all of our samples are clustered at low values of $\tau_0$ suggests that an unknown, extrinsic defect may be polluting the entire sample set. The purity of the precursor chemicals used here is 99.99 at. % for both thermal evaporation and ALD. These are low purity levels for semiconductor synthesis. An impurity with concentration 0.01 at. % in the precursors could be present at concentrations up to approximately $10^{18}$ cm$^{-3}$ in the films. If such impurity forms a recombination-active defect then this would dominate the minority-carrier recombination and would frustrate attempts to increase $\tau_0$ by controlling intrinsic defects. We suggest that synthesizing SnS thin films using semiconductor-grade chemical precursors and in a suitably clean growth environment may yield films with significantly higher $\tau_0$.

We are making freely available software that implements the analysis described in **Section 4d-g** in an easy-to-use graphical format in the Matlab environment.[42] It also allows global fits to multiple data sets with flexible choices of fixed and free parameters. Here we use this global fitting routine to fit the model to multiple data sets collected from individual samples. However, it could also be used to fit the model to data collected from separate samples. This would enable a more flexible form of hypothesis testing akin to Bayesian inference. For example, it would be possible to test the hypothesis that certain post-growth oxidation procedures affect *S* but not $\tau_0$ by comparing goodness-of-fit metrics.

Minority-carrier lifetime is one of the most important figures of merit for a solar cell absorber and is no less important than frequently discussed metrics such as the band gap, optical absorption and phase stability. However, unlike most other metrics, minority-carrier lifetime depends strongly on materials processing, and can vary by orders of magnitude for a single material composition (*c.f.* **Figure 1**). Lifetime metrology is therefore key to developing solar cell technology, both for established materials and for early-stage materials alike. The difficulty of lifetime metrology on thin films compared to wafer-based materials should not prevent research targeted at improving $\tau_0$ in thin films. Lifetime metrology can be used to assess the impact of materials processing more quickly and with greater accuracy than the fabrication and test of complete solar cell devices. Lifetime metrology can also be used to screen new materials as solar cell absorbers. The defining characteristic of $MAPbX_3$ is the large $\tau_0$ that can be achieved with much lower processing temperatures and source purities than are needed for more typical semiconductors. An emphasis on lifetime metrology would accelerate the search for a Goldilocks material: a solar cell absorber with large $\tau_0$, an appropriate band gap and strong optical absorption that is also phase-stable, non-toxic, and can be manufactured at low cost.[1]

**Acknowledgments**

We thank R. E Brandt and for discussions. This work is supported by the U.S. Department of Energy through the SunShot Initiative under contract DE-EE0005329, and by Robert Bosch LLC through the Bosch Energy Research Network under grant 02.20.MC11. R. Jaramillo, B. K. Ofori-Okai, V. Steinmann, and K. Hartman acknowledge the support of a DOE EERE Postdoctoral Research Award, the NSF GRFP, the Alexander von Humboldt foundation, and an Intel PhD Fellowship, respectively. This work is supported by the US Department of Energy, Basic Energy Science, Materials Science and Engineering Division. This research was funded by the Global Climate and Energy Project. This work was partially supported by the Department of Energy Grant No. DE-FG02-00ER15087 and the National Science Foundation Grant No. CHE-1111557.

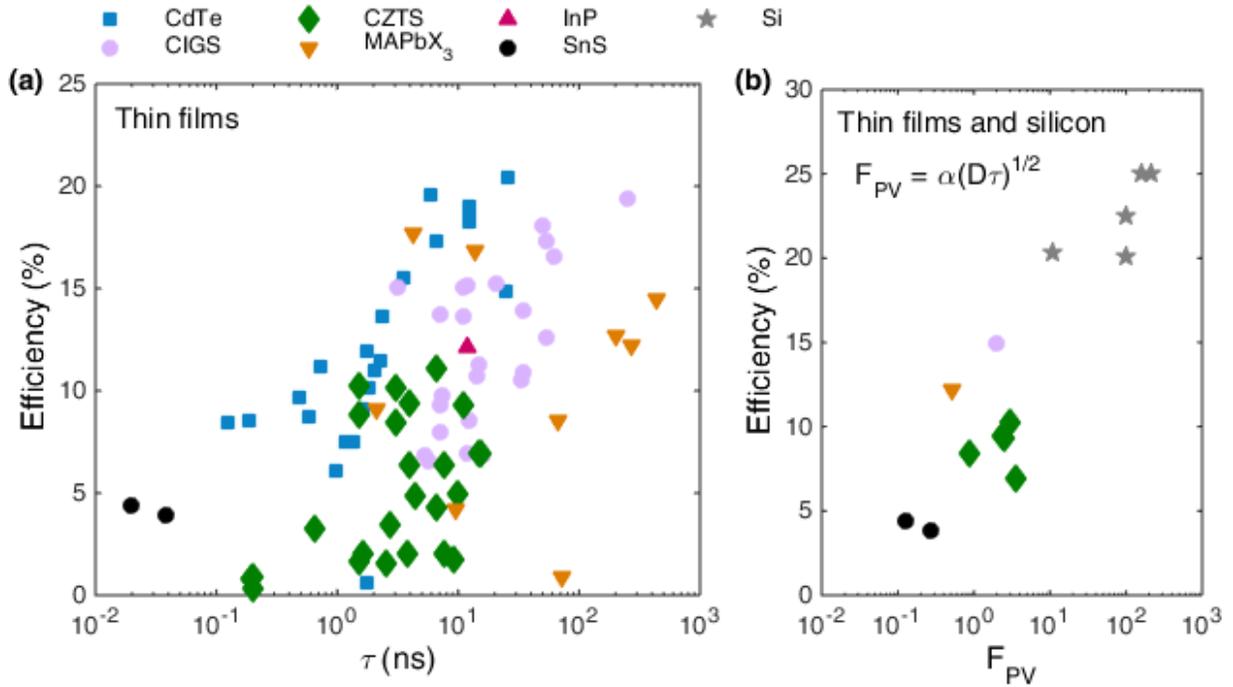

**Figure 1:** Minority carrier lifetime ($\tau$) and a solar cell figure of merit *vs* device efficiency. Efficiency values are for AM1.5 insolation. (a) Minority-carrier lifetime *vs* device efficiency for polycrystalline thin film absorbers. (b) Figure of merit ($F_{PV}$) *vs* device efficiency for thin films and crystalline silicon wafers. In (a) we only show data for which both $\tau$ and device measurements were performed on samples that were synthesized in the same laboratory and using as close to the same procedure as is reasonably possible; in (b) we additionally require for thin films that $D$ and $\alpha$ are reported on comparable samples. For silicon we use tabulated values for $D$ and $\alpha$. For SnS, the efficiencies are as reported in refs. 2 and 3. For other materials the sources of data are: CdTe, refs. 43-46; CIGS, refs. 47-50; CZTS, refs. 50-54; MAPbX3, refs. 28–32; InP, refs.55 and 56; Si, refs. 57–60

| *Sample name & number* | *Substrate* | *Growth* | *H₂S annealing* | *Surface treatment* |
|---|---|---|---|---|
| TE1 (SnS141016p) | Fused quartz | TE | None | < 15 sec air exposure |
| TE2 (SnS140901e) | Fused quartz | TE | None | 24 hrs air exposure |
| TE3 (SnS140901a) | Fused quartz | TE | Short anneal (28 Torr, 4% H2S in $N_2$, 400 °C, 1 hr) | 24 hrs air exposure |
| TE4 (SnS140901d) | Fused quartz | TE | Long anneal (80 Torr, 2% $H_2S$ + 2% $H_2$ in $N_2$, 450 °C, 3 hrs, with 2 hr linear controlled cooldown) | 24 hrs air exposure |
| TE5 (SnS141016n) | Fused quartz | TE | Short anneal | Zn(O,S):N (1 super cycle, approximately 3 nm thick) |
| TE6 (SnS141016o) | Fused quartz | TE | Short anneal | UV/ozone exposure (5 s) |
| TE7 (SnS140901h) | Fused quartz | TE | None | O2 plasma (10 W, 0.5 Torr, 12 sec) |
| TE8 (SnS140901f) | Fused quartz | TE | None | $H_2O_2$ exposure (5 cycles of 2.1 Torr s, for a total of 10.5 Torr s) |
| TE9 (SnS140901b) | Fused quartz | TE | Short anneal | $H_2O_2$ exposure |
| ALD1 (Solar093B) | Fused quartz | ALD | None | $H_2O_2$ exposure |
| ALD2 (Solar093D) | Fused quartz | ALD | Short anneal (10 Torr, 99.5% $H_2S$, 400 °C, 1 hr) | $H_2O_2$ exposure |

**Table 1:** SnS thin film sample set used for lifetime measurements. TE = thermal evaporation, ALD = atomic layer deposition. Previous results suggest that H₂S annealing improves the bulk minority-carrier lifetime, and that oxidation of the surface lowers the SRV. For sample TE5, the Zn(O,S):N composition is the same as used in previously published devices.[2,3,13]

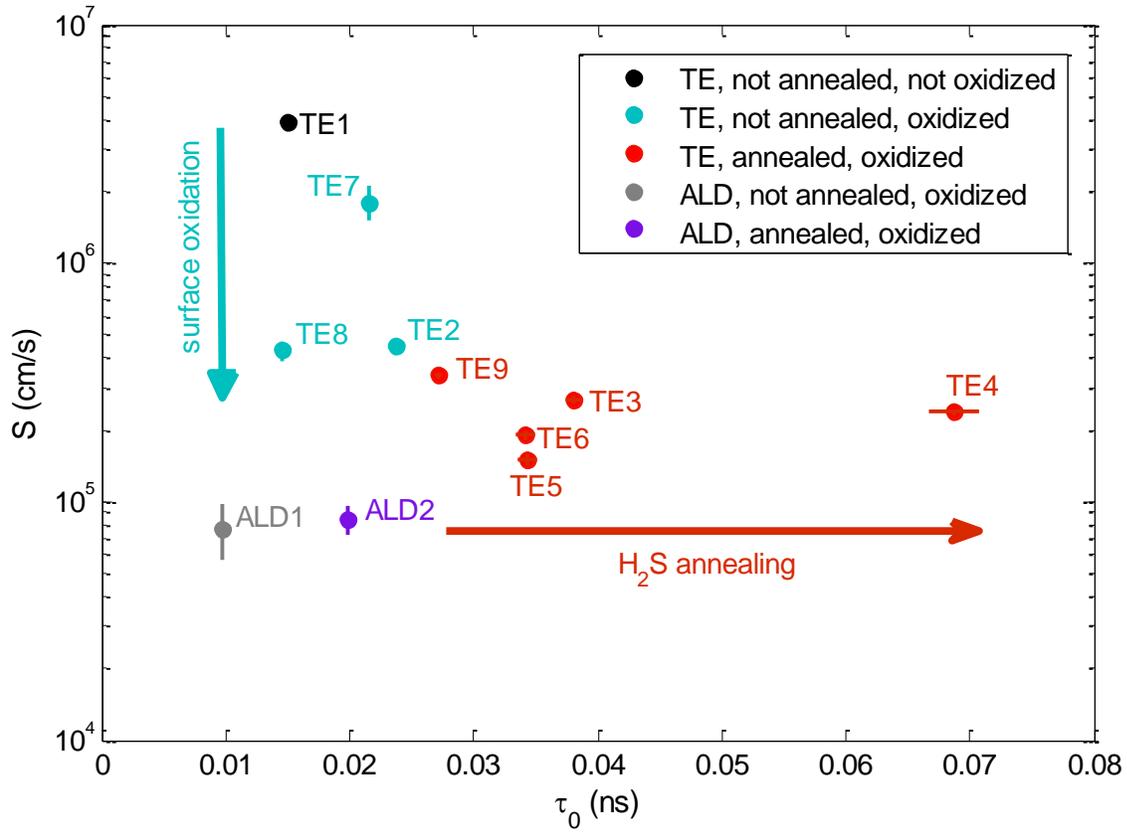

**Figure 2:** Minority-carrier lifetime ($\tau_0$) and SRV ($S$) for the full sample set described in **Table 1**. Error bars indicate the 95% confidence intervals of the global fits. The arrows summarize the trends in the data, and confirm our hypotheses based on previous studies of completed solar cells.

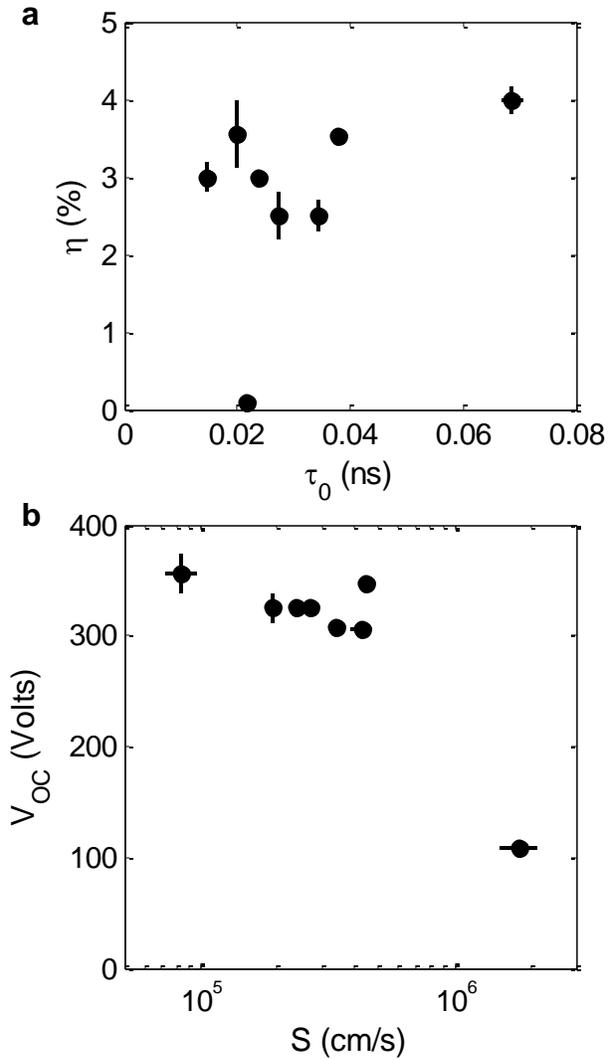

**Figure 3:** Dependence of SnS solar cell device performance on minority-carrier recombination. (a) $\eta$ compared to $\tau_0$. Due to the relatively small range of $\tau_0$ across our sample set, the effect of changing $\tau_0$ is easily masked by competing factors. (b) $V_{OC}$ compared to $S$. The decreasing trend in $V_{OC}$ with increasing $S$ is consistent with the prominent contribution of interface recombination to the total recombination current near open-circuit voltage.

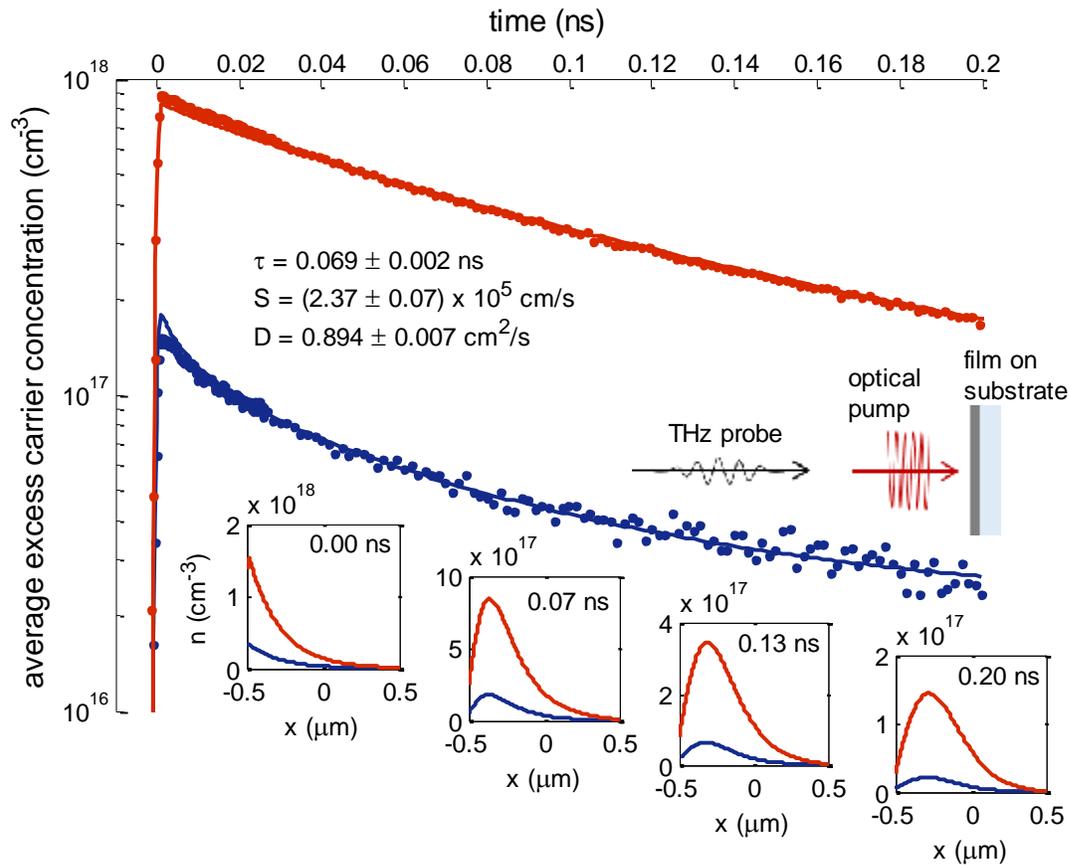

**Figure 4:** TPC data and fits for a single sample (TE4). The sample was measured with 800 nm (red) and 400 nm (blue) pump wavelengths. Points are data, and lines are the fitted model. Both data sets were fit simultaneously to determine $\tau$, $S$, and $D$. The main axis shows $\bar{n}(t)$, the excess minority-carrier concentration averaged through the sample thickness. The insets show $n(x)$, the spatial distribution of the excess-carrier concentration, at four points in time according to the fitted model. The schematic illustration shows the film-on-substrate sample structure, the tunable optical pump pulse, and the THz probe pulse.

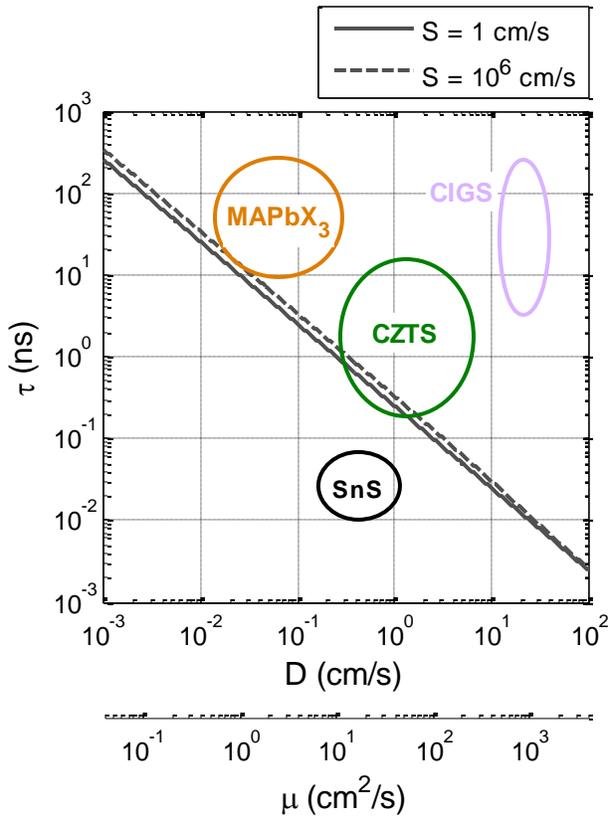

**Figure 5:** Parameter spaces on the $D$-$\tau$ plane within which the spatial distribution of excess-carriers can and cannot be ignored when fitting TPC data for fixed film thickness $d = 1$ μm. The grey lines satisfy the condition $\tau_{eff} = t_d/4\pi^2$ for discrete values of $S$. The second x-axis shows mobility (μ) at 293 K. For $\tau_{eff} \gg t_d/4\pi^2$ (upper right) the fundamental mode is easily measured, and the spatial distribution of excess-carriers does not affect the estimation of $\tau_{eff}$. For $\tau_{eff} \ll t_d/4\pi^2$ (lower left) the data is not well-characterized by a single exponential decay, and the spatial distribution of excess-carriers must be considered in order to estimate $\tau_{eff}$. The circles indicate typical parameter spaces for different thin film materials, with the same color scheme as **Figure 1**.

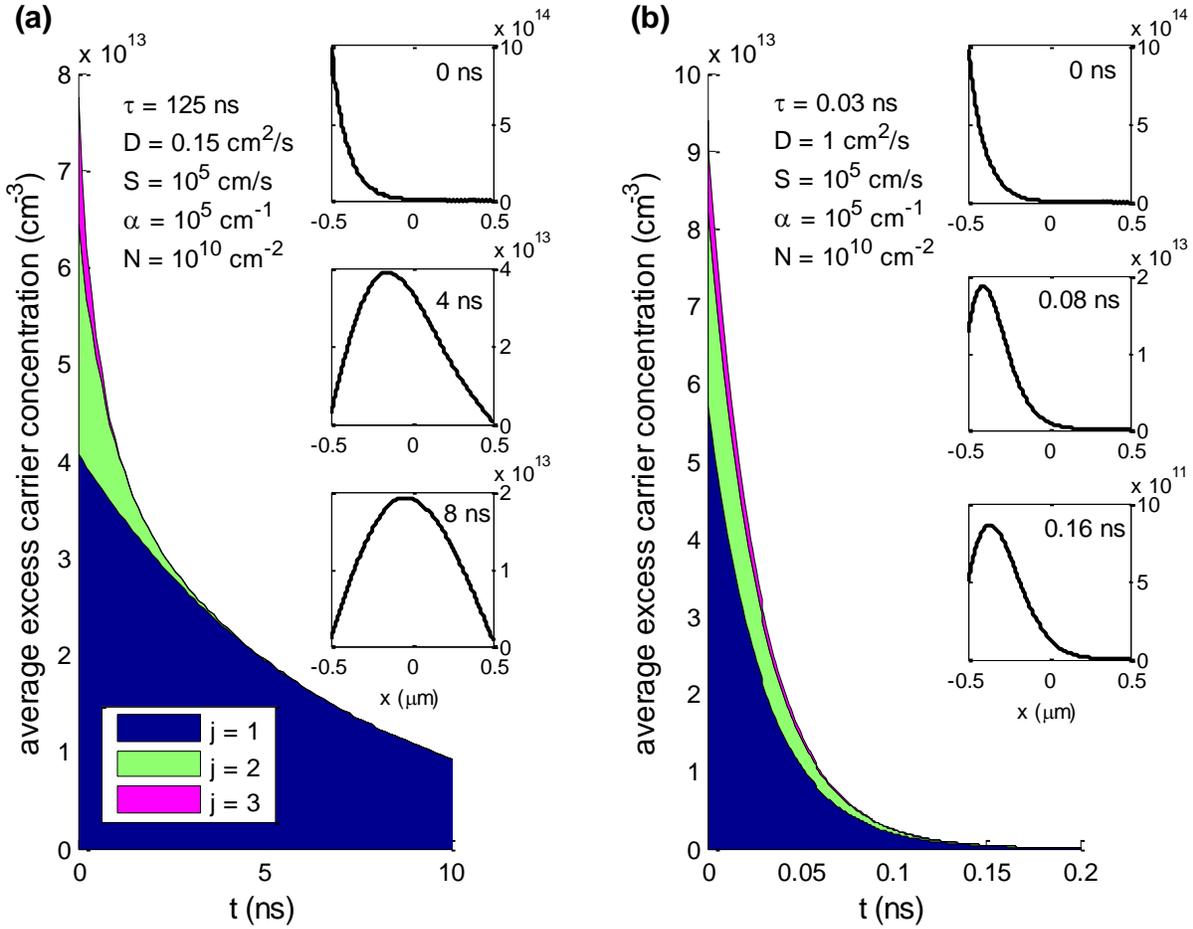

**Figure 6:** Decay of spatial frequencies in a TPC experiment, calculated using the solution developed in ref. 16 and described by Eq. 4. The main panels show stacked plots of the first three spatial frequencies (j = 1, 2, 3) that contribute to $\bar{n}(t)$. The insets show the spatial distribution $n(x)$ at discrete times. (a) Material with long minority-carrier lifetime and small diffusivity, typical of MAPbX$_3$. (b) Material with short minority-carrier lifetime and larger diffusivity, typical of SnS. For both simulations S = $10^5$ cm/s, the optical absorption coefficient ($\alpha$) is $10^5$ cm$^{-1}$, and the film thickness is 1 µm. The pump is a delta function pulse of light at $t = 0$ with fluence $N_0 = 10^{10}$ cm$^{-2}$. The solution is linear in $N_0$, so that a different choice of $N_0$ would simply scale the vertical axes without affecting the dynamics or the spatial distributions.

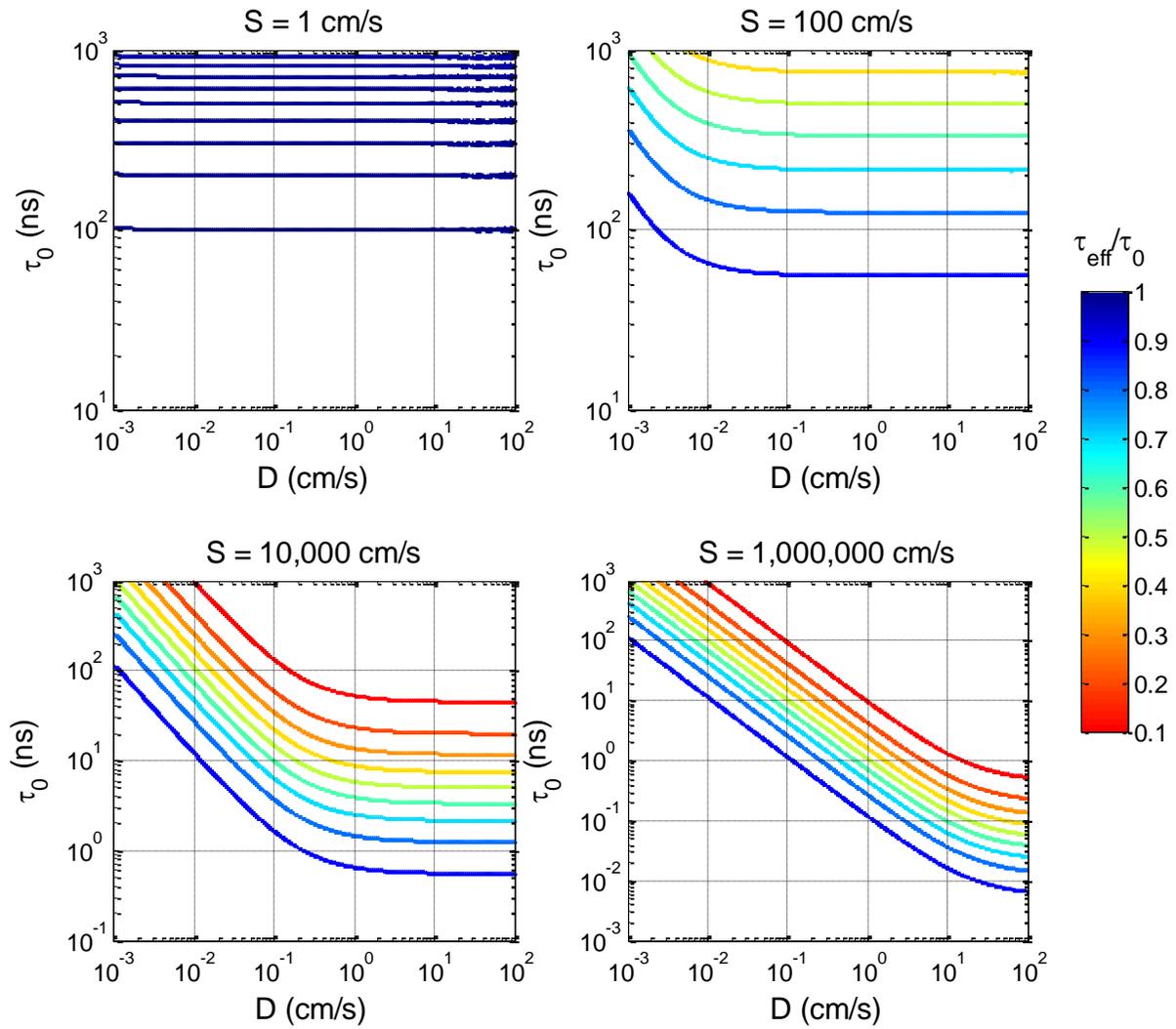

**Figure 7**: Ratio of effective lifetime $\tau_{eff}$ to the bulk minority-carrier lifetime $\tau_0$ as a function of $D$ and $\tau$ for discrete values of S and film thickness $d = 1$ μm. For $\tau_{eff}/\tau_0 \approx 1$ the minority-carrier lifetime can be estimated from the effective lifetime. For $\tau_{eff}/\tau_0 < 1$ the effective lifetime is affected by both bulk and surface recombination.

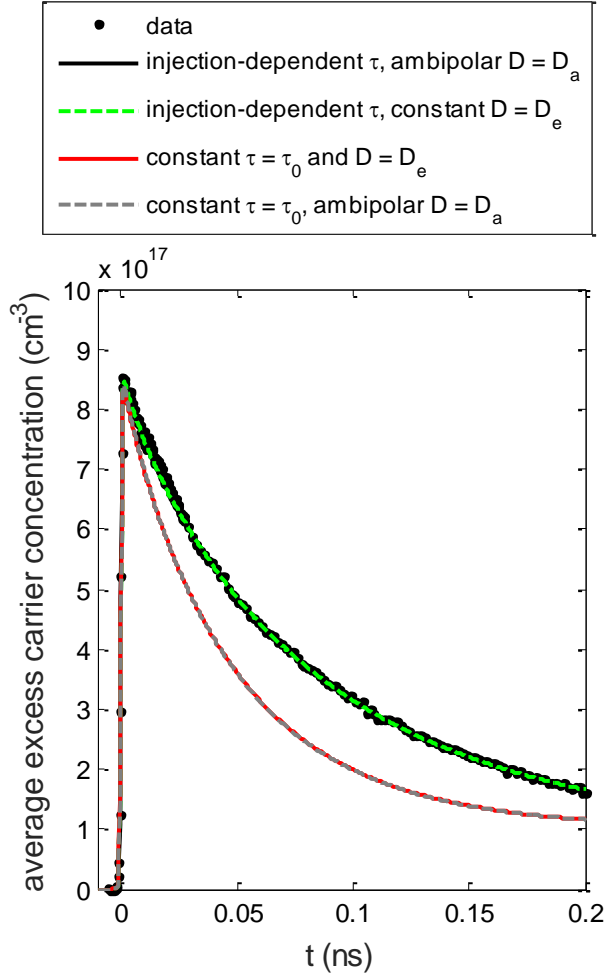

**Figure 8**: Effects of high-injection and ambipolar diffusion on the thickness-averaged excess minority-carrier concentration $\bar{n}(t)$. Solid lines show the solutions to the nonlinear diffusion model, and points are experimental data. The data are for sample TE4 with 800 nm pump wavelength, the same as presented in **Figure 4**. For this sample $p_0 = 8.07 \times 10^{16}$ cm$^{-3}$, and for this dataset the pump fluence $N_0 = 1.01 \times 10^{14}$ cm$^{-2}$. Black solid line: model with injection-dependent $\tau$ and ambipolar $D = D_a$, as described in Eq. 4-5 and used to generate the results reported here. Green dashed line: model with injection-dependent $\tau$ and constant $D = D_e$. Red solid line: model with constant $\tau = \tau_0$ and $D = D_e$. Grey dashed line: model with constant $\tau = \tau_0$ and ambipolar $D = D_a$.

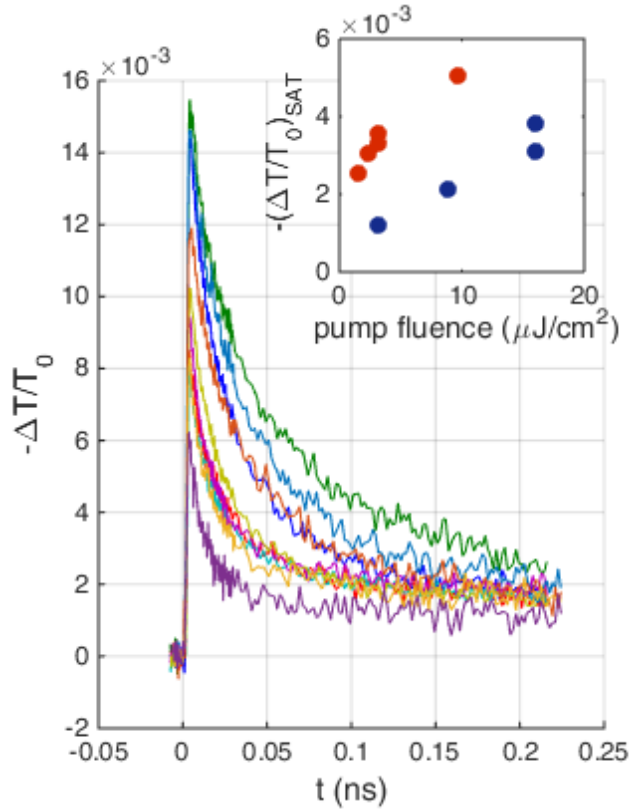

**Figure 9**: Saturation of $\Delta T/T_0$ at non-zero values $(\Delta T/T_0)_{SAT}$ at long times. The main panel shows $\Delta T/T_0$ measured for a 400 nm, 11 µJ/cm² pump. The different colored lines indicate different samples. The transient decay reflects the recombination of excess-carriers, and the size of the peak at short times depends on the sample mobility (*c.f.* **Eq. 6**). The saturation at long times is of unknown origin. (Inset) $(\Delta T/T_0)_{SAT}$ as a function of pump fluence and wavelength for an individual sample (sample TE3). The blue and red points indicated 400 and 800 nm pump, respectively.

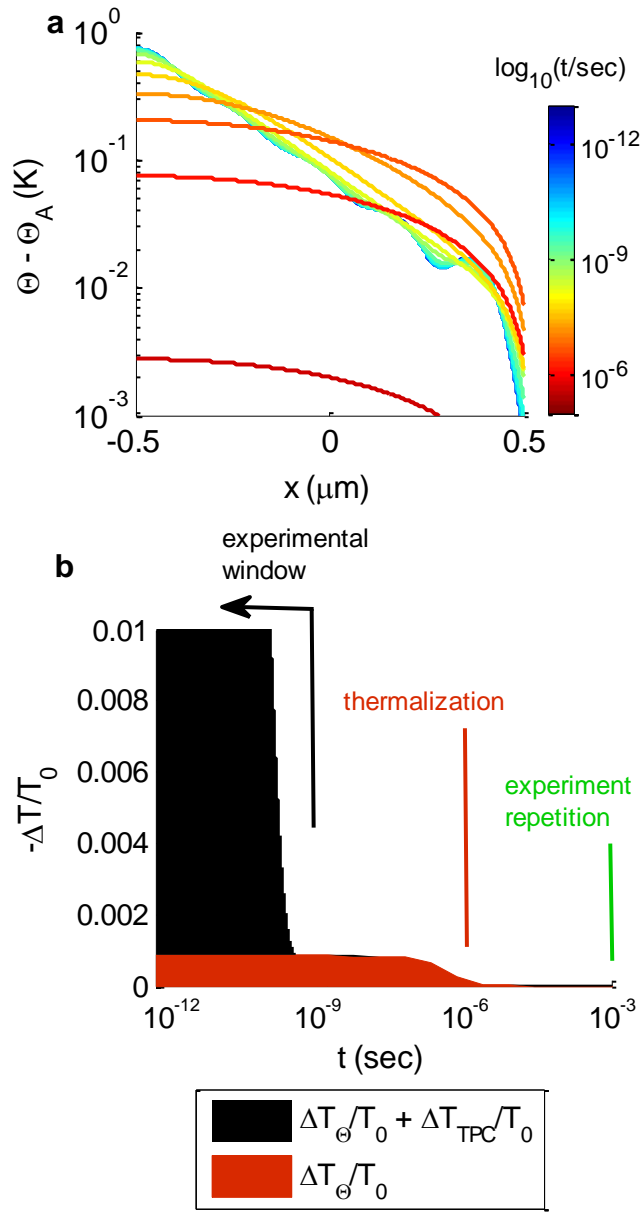

**Figure 10**: Effect of transient sample heating on measured $\Delta T/T_0$. **(a)** Temperature profile $\Theta(x, t)$ though a 1 mm thick SnS film on fused quartz as a function of time ($t$) after an instantaneous laser pulse that is incident on the surface at $x = -0.5$ μm. The simulation is described in the text and the laser pulse has wavelength 800 nm and total energy density 25 μJ cm$^{-2}$. $\Theta(x, t)$ decays quickly to zero for $t \gg t_\Theta = 1$ μs. The oscillations at short times are artifacts due to numerical truncation. **(b)** We average the temperature rise in (a) through the film thickness and convert it to a thermally-induced rise $\Delta T_\Theta/T_0$ (black filled area) using an independent measurement of THz transmission as a function of temperature. For illustration we add to this thermal transient the photoconductivity signal $\Delta T_{TPC}/T_0$ that we calculate from the model using parameters typical for sample TE3 and $f = 0$. The y-axis is truncated well below the maximum in $\Delta T_{TPC}/T_0$ in order to emphasize the long transient $\Delta T_\Theta/T_0$, which is similar in magnitude to the measured $(\Delta T/T_0)_{SAT}$.